# MultiWalk: A Framework to Generate Node Embeddings Based on an Ensemble of Walk Methods

Kaléu Delphino
kdelphino3@gatech.edu

*Abstract*—Graph embeddings are low dimensional representations of nodes, edges or whole graphs. Such representations allow for data in a network format to be used along with machine learning models for a variety of tasks (e.g., node classification), where using a similarity matrix would be impractical. In recent years, many methods for graph embedding generation have been created based on the idea of random walks. We propose MultiWalk, a framework that uses an ensemble of these methods to generate the embeddings. Our experiments show that the proposed framework, using an ensemble composed of two state-of-the-art methods, can generate embeddings that perform better in classification tasks than each method in isolation.

## I. INTRODUCTION

Many datasets of research interest may be modeled as a graph. Examples are social networks [2], the Web [3], protein-protein interaction [4] and airport traffic [11]. Although graphs are a natural representation in those cases, there is not a straight way to extract the information in them and apply it to machine learning models. A pairwise similarity matrix such as the adjacency matrix, while providing all the information in numerical form, can be too big to be dealt with. To circumvent this limitation, a common technique is to transform the graph (in its entirety or only its nodes or edges) into embeddings. Embeddings are projections of graph data into a $k$-dimensional space. Researchers usually concentrate on building algorithms that can learn embeddings from graphs in an unsupervised way to avoid time-consuming tasks related to feature engineering.

In order to be of practical use, node embeddings need to have a much lower dimension than the number $n$ of nodes in the graph (i.e., $k \ll n$). Therefore, the challenge is to generate embeddings that are at the same time compact enough and still keep useful information from the original graph.

One way to achieve that is by making embeddings close between each other (under some distance metric such as Euclidean) if and only if their correspondent nodes are similar in the graph (in some well-defined similarity metric). Based on that idea, Perozzi et al. [5] proposed *DeepWalk*, a method where a number of random walks are performed in the graph. A random walk is a sequence of adjacent nodes taken randomly, such as *a, b, a, c, d* in Figure 1. Those walks are used as training data in a language model that is capable of generating embedding for words given sentences. Following this pioneering work, other similar methods have been created [6]. Many of them repeat the same steps as in *DeepWalk,* but use a modified version of the pure random walk hoping it is able to better capture the graph information.

Given the current profusion of such methods, we propose *MultiWalk*, a framework where different walk methods can be combined to generate high-quality embeddings. More specifically, our framework gathers walks from different sources and then apply them to a single language model. We start by giving a brief literature review on the subject in Section 2. In Section 3, we describe our framework. In Section 4, we present experiments with the framework along with the results. Finally, a short conclusion is given in Section 5.

## II. RELATED WORK

First works in graph embedding used methods based on matrix factorization [ref]. The more recent surge on random walk-based methods follows the good results of *DeepWalk* [5]. *DeepWalk* performs random walks in a graph and use the sequences of nodes as sentences in the *SkipGram* model [7], a language model originally built to generate embeddings for words in a text corpus. In that setting, nodes are similar if they appear in the same walk (limited by a sliding window of fixed length). Not only *DeepWalk* can generate embeddings of high quality in a variety of scenarios, it is also highly scalable, allowing graphs of millions of nodes and edges to be processed in regular domestic computers [ref]. In an import theoretical result, Qiu et al. [8] proved that *DeepWalk* and similar methods are implicitly factorizing a matrix.

Many variants of *DeepWalk* have since been proposed [6]. *node2vec* [21] is a variation of *DeepWalk* where the random walks can be biased to have characteristics of deep and breadth-first search. In [13], edge weights are considered to choose the next node in a walk. In [14], random walks are applied in heterogeneous graphs where nodes can be different in nature. For example, in the same graph, different nodes may represent users, questions and answers. In [15], the random walks skip nodes to capture high-order relationships. In [16], node attribute vectors are incorporated by creating another graph where each node has an edge to nodes with similar vectors. This new graph is connected to the original graph, forming a new two-layered graph. The random walks are then performed in the final graph.

As the nodes that share the same vicinity tend to appear in the same walks, the embeddings generated by methods that solely consider them tend to be clustered in different communities of nodes that are close in the graph, a concept also referred to as homophily [9, 10]. Still, nodes far apart from each other can be considered similar if they hold the same structural properties. For example, in some problems, a node acting as a hub (i.e., a node with a relative high degree compared to its neighbors) should be considered similar to another hub in a distance part of the network. To address this issue, some effort has been directed to create methods that generate embeddings that can capture the structural characteristics under the random-walk framework. The best example of such methods is *struc2vec* [11]. In *struc2vec*, a multi-layered graph is created with the same nodes as the ones in the original graph, but with edges favoring connections between structurally similar nodes. The random walks are then performed in this new graph. Another attempt to capture graph structural features via random walks is using anonymous random walks [12]. Other works try to find a balance in capturing structural and proximity properties [17, 18, 21].

Our proposed framework is based on the concept of ensemble learning [19], where two or more models are taken together to produce a better result than using the models separately. Ensembles have a long history in Machine Learning (for an early work, see [20]). In the context of graph embedding generation, Chen and Papalexakis [1] proposed *TenSemble2vec*, an ensemble-based technique that learns a shared embedding among many previously generated embeddings.

We highlight that our framework is not based on the usual concept of ensemble that summarizes the *outputs* of different full-fledged models, but it considers *walks* from different methods before they are passed to a single language model. To the best of our knowledge, this idea has not been explored in the literature.

### III. THE MULTIWALK FRAMEWORK

Consider the graph $G = (V, E)$ where $V$ is the set of nodes (vertices) of $G$ and $E \subseteq (V \times V)$ is the set of edges. A walk generator is a function that produces finite sequences of nodes in $G$. Notice that, despite its name, walk generators are allowed to output any arbitrary sequence of nodes, and there is no need of a node to be followed by one that is adjacent to it. The name "walk" is used here to imply that, while sequences may be totally arbitrary, we expect that the function will follow some concept of adjacency.

---

**Algorithm 1** MultiWalk

**Input:**
  Graph $G$
  List of walk generators $M$
  List of number of walks per generator $N$
  Set of SkipGram arguments $S$
**Output:** Matrix of node embeddings $\phi$

1: **procedure** GENERATEEMBEDDINGS($G$, $M$, $N$, $S$)
2:   Initializate $W$ as an empty list
3:   **for each** $v$ in $G$ **do**
4:     **for each** $m$ in $M$ **do**
5:       **for** $i = 0$ to $N[m]$ **do**
6:         append a walk generated by $m$ to $W$
7:   $\phi = \text{SkipGram}(W, S)$

---

A specification for *MultiWalk* is given by Algorithm 1. A list of walk generators along with the number of walks that each generator should produce for each node is passed to the procedure. Then, a number of walks are created starting from each node. The total of walks per node is fixed since it is comprised of the sum of all fixed numbers of walks per method. Finally, all the generated walks are passed to the *SkipGram* language model to get the corresponding node embeddings.

Two important points should be noted. First, the length of the walks can be different for each walk generator. This allows for more flexibility when applying multiple generators. Second, the algorithm guarantees that each method will provide their share of walks starting at each node.

### IV. EXPERIMENTS

We performed experiments to test the *MultiWalk* framework using different blends of walks from two walk generators: the random walk generator as proposed in *DeepWalk*, and the walk generator used in the *struc2vec*. We also run the same experiments using both methods in isolation (i.e., outside the framework) and compared the results.

The reason we chose these two methods is that they are at the extreme ends of generating embeddings that preserve homophily and structural similarity. The random walks of *DeepWalk* will always have nodes and their neighbors in them. In the case of struc2vec, a single walk (performed in a modified graph) will contain nodes that are structurally similar. In joining these two very dissimilar methods, we expect to find a "sweet spot" capable of providing better embeddings than if each method was used separately.

We chose node classification using one-vs-rest logistic regression as the task to assess performance. In this setting, we aim to measure how well given node embeddings can perform in predicting unseen previously labeled nodes by only using a portion of the dataset to generate them. For each dataset, we randomly split the data into training and test sets in an 80-20 proportion, and the experiments were repeated 10 times.

The following subsections share the details of the experiments.

#### A. Code and Computational Setup

All the code for the experiments was programmed using Python 3 and run on a Windows 64-bits machine. We implemented the code to generate *DeepWalk* random walks, and the *struc2vec* walks were generated by the reference code provided by its authors.

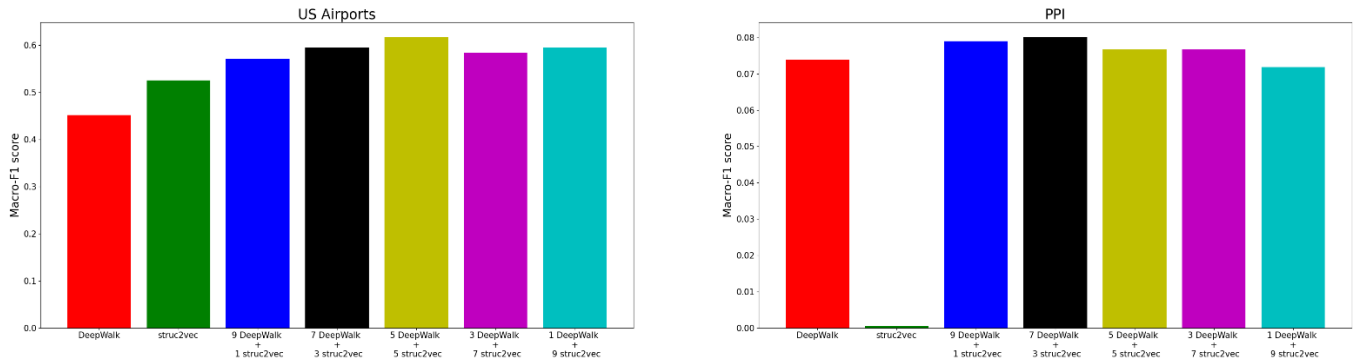

*Figure 1 - Average macro-F1 score for node classification using embeddings generated by different methods after 10 rounds. Scores are based on performance in 20% of the data. Methods based on MultiWalk have names in the format "number of DeepWalk walks + number of struc2vec walks". Left: results in the US Airports dataset. Right: results in the PPI dataset.*

### B. Datasets

We performed the experiments in the following datasets, which have already been used to assess the quality of node embeddings in other works [8, 11]:

- *United States air-traffic network (US Airports)*: dataset comprised of a graph that connects airports in the US that had at least one direct flight between them in the period of January to October 2016. There is a total of 1,190 nodes and 13,599 edges in the graph. Each airport is labeled with one of four possible different categories that are based on the number of flights that passed by them (departures and arrivals).

- *Protein-Protein Interactions (PPI)*: according to [21], this is a graph of a subset of the PPI network [22] for *Homo sapiens*. It has 3,890 nodes and 76,584 edges. Each node is labeled by one or more labels from a set of 40 different labels.

### C. Parameters and other implementation details

The *SkipGram* parameters, shared by all the tested methods, were dimension $d = 128$, window size $w = 10$, number of epochs $i = 5$ and number of negative samples $\pi = 5$. For both *DeepWalk* and *struc2vec* original versions, we used number of walks $n = 10$ and walk length $l = 80$.

For the methods used within the *MultiWalk* framework, we created five variations, each one based on a different pair ($num\_walks_{DW}$, $num\_walks_{s2v}$), where $num\_walks_{DW}$ is the number of walks sampled by a regular random walk and $num\_walks_{s2v}$ is the number of walks sampled by the *struc2vec* method. The specific pairs used where (9, 1), (7, 3), (5, 5), (3, 7) and (1, 9). All the pairs were chosen to make the number of walks sum up to 10, the same number of walks used in the experiments with the models outside the framework.

In order to obtain more reliable results about the difference of performance between the models, all methods sampled their walks from the same pools of precomputed walks, one pool for regular random walks and the other for *struc2vec* walks. Each pool had 30 precomputed walks for each starting node. In this fashion, we expect all the *MultiWalk* methods to have similar walks as the non-*MultiWalk* methods during the training, and therefore differences in performance are more related to the effect of mixing different types of walks than to the specific walks used.

### D. Results

The results of the experiments can be seen in Figure 1. In both datasets, using our proposed framework to mix walks from both methods performed better than using the models independently. In the US Airports dataset, the best model was the one who used the same number of walks from each generator. In the PPI dataset, the best model was the one who used 7 *DeepWalk* walks and 3 *struc2vec* walks. An interesting observation is that, while *struc2vec* performed very poorly in the PPI dataset, it could still boost overall performance when used in conjunction with *DeepWalk*.

## V. CONCLUSION

In this paper, we proposed a framework for node embedding generation that uses an ensemble of walk generator methods. Experiments were able to demonstrate that this approach can outperform state-of-the-art models that are based on a single walk generator, by aggregating the existent strengths in each of the methods.

## REFERENCES

[1] J. Chen and E. Papalexakis. Ensemble Node Embeddings using Tensor Decomposition: A Case-Study on DeepWalk. arXiv preprint arXiv:2008.07672, 2020

[2] M. Newman, D. Watts and S. Strogatz. Random graph models of social networks. In *Proceedings of the National Academy of Sciences*, 99 (suppl 1), 2566–2572, 2002.

[3] R. Kumar, P. Raghavan, S. Rajagopalan, D. Sivakumar, A. Tompkins, and E. Upfal. The Web as a Graph. In *Proceedings of the Nineteenth ACM SIGMOD-SIGACT-SIGART Symposium on Principles of Database Systems*, pages 1–10. 2000.

[4] F. Yang, K. Fan, D. Song, and H. Lin. Graph-based prediction of Protein-protein interactions with attributed signed graph embedding. *BMC Bioinformatics*, 21, 323, 2020.

[5] B. Perozzi, R. Al-Rfou, and S. Skiena. 2014. DeepWalk: Online Learning of Social Representations. In *ACM SIGKDD*, 2014

[6] H. Cai, V. Zheng and K. Chang, A Comprehensive Survey of Graph Embedding: Problems, Techniques, and Applications. In *IEEE Transactions on Knowledge and Data Engineering*, 30(9):1616-1637, 2018.

[7] T. Mikolov, K. Chen, G. Corrado, and Jeffrey Dean. Efficient Estimation of Word Representations in Vector Space. In *ICLR*, 2013.

[8] J. Qiu, Y. Dong, H. Ma, J. Li, K. Wang, and J. Tang. Network Embedding as Matrix Factorization: Unifying DeepWalk, LINE, PTE, and node2vec. In *WSDM*, 2018.

[9] Birds of a Feather: Homophily in Social Networks. M. McPherson, L. Smith-Lovin, J, Cook. *Annual Review of Sociology* 27, 1, 415-444, 2001.

[10] D. Easley and J. Kleinberg. Networks in Their Surrounding Contexts. In *Networks, Crowds, and Markets: Reasoning about a Highly Connected World*. Cambridge University Press, 2010, pp. 85-118.

[11] L. Ribeiro, P. Saverese, and D. Figueiredo. struc2vec: Learning node representations from structural identity. In *KDD*, 2017.

[12] S. Ivanov, and E. Burnaev. Anonymous Walk Embeddings. In *ICML*, 2018.

[13] Z. Jin, R. Liu, Q. Li, D. Zeng, Y. Zhan, and L. Wang. Predicting user's multi-interests with network embedding in health-related topics. In *IJCNN*, 2016.

[14] H. Fang, F. Wu, Z. Zhao, X. Duan, Y. Zhuang, and M. Ester. Community-based question answering via heterogeneous social network learning. In *AAAI*, 2016.

[15] B. Perozzi, V. Kulkarni, H. Chen, and S. Skiena. Don't Walk, Skip! Online Learning of Multi-Scale Network Embeddings. In *ASONAM*, 2017.

[16] S. Bandyopadhyay, A. Biswas, H. Kara, and N. Musti. A Multilayered Informative Random Walk for Attributed Social Network Embedding. In *European Conference on Artificial Intelligence*, 2020

[17] S. Cao, W. Lu, and Q. Xu. GraRep: Learning Graph Representations with Global Structural Information. In *Proceedings of the 24th ACM International on Conference on Information and Knowledge Management*, 2015.

[18] B. Shi, C. Zhou, H. Qiu, X. Xu, and J. Liu. Unifying Structural Proximity and Equivalence for Network Embedding. In *IEEE Access*, vol. 7, pp. 106124-106138, 2019.

[19] J. Friedman, T. Hastie, and R. Tibshirani. *The Elements of Statistical Learning*. Springer, 2001.

[20] L. Hansen and P. Salamon. Neural Network Ensembles. In *IEEE Transactions on Pattern Analysis and Machine Intelligence*, vol. 12, no. 10, pp. 993-1001, Oct. 1990.

[21] A. Grover and J. Leskovec. node2vec: Scalable Feature Learning for Networks. In *KDD*, 2016.

[22] B. Breitkreutz, C. Stark, T. Reguly, L. Boucher, A. Breitkreutz, M. Livstone, R. Oughtred, D. H. Lackner, J. Bähler, V. Wood, et al. The BioGRID interaction database. *Nucleic Acids Research*, 36:D637–D640, 2008.